%% file: template.tex
\documentclass{article}

\usepackage{arxiv}

\usepackage[utf8]{inputenc} 
\usepackage[T1]{fontenc}    
\usepackage{hyperref}       
\usepackage{url}            
\usepackage{booktabs}       
\usepackage{amsfonts}       
\usepackage{nicefrac}       
\usepackage{microtype}      
\usepackage{caption}
\usepackage{amsmath}
\usepackage{amsthm}
\usepackage{bbold}
\usepackage{siunitx}
\usepackage{longtable}
\usepackage{array}
\usepackage{multirow}
\usepackage{wrapfig}
\usepackage{float}
\usepackage{colortbl}
\usepackage{pdflscape}
\usepackage{tabu}
\usepackage{threeparttable}
\usepackage{threeparttablex}
\usepackage[normalem]{ulem}
\usepackage{makecell}
\usepackage{xcolor}
\usepackage[round]{natbib}
\usepackage{enumerate}
\usepackage[shortlabels]{enumitem}
\usepackage{pgfplots}
\usepackage{hyperref}

\usepackage{ulem} 
\usepackage{tabularray} 
\usepackage{float} 
\usepackage{graphicx} 
\usepackage{rotating} 
\usepackage[normalem]{ulem} 
\UseTblrLibrary{booktabs} 
\UseTblrLibrary{siunitx} 
  \NewTableCommand{\tinytableDefineColor}[3]{\definecolor{#1}{#2}{#3}}

\title{Generalised raking and stabilised weights for regression modelling in two-phase samples}

\author{
Tong Chen \\
Clinical Epidemiology and Biostatistics Unit\\
Department of Paediatrics, University of Melbourne\\
Clinical Epidemiology and Biostatistics Unit\\  
Murdoch Children’s Research Institute\\
Email: tong.chen8@unimelb.edu.au\\
\AND
Joshua Slone \\
Department of Biostatistics\\
Vanderbilt University \\
\AND
Gustavo Amorim\\
Department of Biostatistics\\
Vanderbilt University \\
\AND
Pamela A. Shaw \\
Biostatistics Division\\ 
Kaiser Permanente Washington Health Research Institute \\
\AND
Bryan E. Shepherd \\
Department of Biostatistics\\
Vanderbilt University \\
\And
Thomas Lumley \\
Department of Statistics\\
University of Auckland\\
}
\pgfplotsset{compat=1.18} 
\begin{document}
\maketitle

\begin{abstract}
In regression models fitted to data from complex survey designs, sampling weights often incorporate non-essential variation, inflating variance estimates. Stabilised weights mitigate this issue by adjusting sampling weights to account for variation explained by covariates. In the context of two-phase sampling, we evaluate the performance of optimal stabilised weights and propose combining the stabilised weight estimator with generalised raking, a class of efficient design-based estimators. This combination improves efficiency by reducing unnecessary weight variation and leveraging information from auxiliary variables. We show this combination can be implemented using the standard statistical package that handles two-phase samples and generalised raking. Simulation studies demonstrate that the proposed estimator enhances precision under realistic two-phase designs, though efficiency gains may be limited in highly informative designs. The developed methods were applied to a large multinational two-phase study of Kaposi sarcoma among people living with HIV.
\end{abstract}


\section{Introduction}

When fitting a regression model to data obtained from a complex probability sample, the sampling probabilities are \emph{ignorable} or \emph{non-informative} if sampling is independent of the outcome variable conditional on the covariates \citep{Little1982Models}. If sampling is ignorable for the model of interest, fitting the regression model without applying sampling weights will give consistent estimation and higher precision than using sampling weights \citep{Lumley2017regression}. Even when sampling is not ignorable for the model of interest, it is sometimes possible to add all important design variables as covariates and thus ignore the weights in analysis \citep{Holt1980Regression, Gelman2007Struggles}. This approach, however, is not always feasible: design variables might be unavailable in public-use datasets, or they might be inappropriate for use as covariates.

In cases where ignoring sampling weights entirely is not feasible, and given that these weights often incorporate non-essential variation, there is a clear need to modify the weights to enhance precision. In this paper, we evaluate modifying sampling weights using stabilised weights, which adjust the original sampling weights to ignore only the part of sampling variation explained by covariates. In particular, \cite{robins2000marginal} constructed stabilised weights by taking the ratio of the marginal probability of sampling (ignoring covariates) to the conditional probability of sampling given covariates. This idea has been independently developed on at least three occasions: by \cite{Magee1998improving} and \cite{pfeffermann1999parametric} in the survey literature and \citet{robins2000marginal} in the context of inverse-probability weighting in causal inference. \citet{robins2000marginal} coined the term \emph{stabilised weights}, which \citet{Lumley2017regression} recommended for use in survey contexts as well. However, the practical application of this approach remains limited in published survey analyses, with notable examples including studies by \citet{SKINNER2012} and \citet{Chambers2022weight}.

We focus on two-phase sampling designs \citep{Neyman1938}, which are widely used in health research to select a subsample for additional measurements from an existing cohort or database \citep{breslow2009using, Shepherd2023Multiwave, amorim2025design} due to resource constraints. Two-phase sampling is a cost-effective approach for data collection. The existing cohort or database can act as the phase-1 sample, in which data on outcomes and several covariates are either routinely collected or readily available for all participants. In the second phase, variables of particular interest, often more costly and difficult to measure, are collected or validated only for individuals selected into the phase-2 subsample. 

Formally, suppose the outcome variable $Y$, covariates $Z$, and additional variables $A$ are available or measured for all individuals in a cohort of size $N$ (phase-1 sample). The variables of interest $X$ are subsequently measured for a probability subsample of $n$ individuals (phase-2 sample). Let $i = 1,\dots, N$ index individuals in the phase-1 cohort, and let $R_i$ denote the indicator of inclusion in the phase-2 subsample. For each individual $i$, let $\pi_i=Pr(R_i=1| Y_i, Z_i, A_i)$ denote the sampling probability and $d_i = 1/\pi_i$ denote the design weight. 

We are interested in fitting a generalised linear model with link function $\phi()$
\begin{equation}
\phi(E[Y|X,Z])=\beta_x X+\beta_z Z
\label{the-model}
\end{equation}
to the two-phase sample, where $\beta = (\beta_x, \beta_z)^\top$ denotes the vector of regression coefficients. Our aim is to improve the efficiency of regression parameter estimates by adjusting the design weights $d_i$. Accordingly, we propose combining stabilised weights with generalised raking, also known as calibration \citep{deville1992calibration, Deville1993Generalized}, an efficient design-based estimator \citep{robins1994estimation, lumley2011connections}. Specifically, generalised raking adjusts design weights by minimising the total weight change under a prespecified distance function, while satisfying a calibration constraint that ensures the weighted total of auxiliary variables equals their known population total. In two-phase sampling, generalised raking is preferable to inverse probability weighted (IPW) estimators \citep{Horvitz1952}, as it improves efficiency by incorporating rich phase-1 information.

Two-phase sampling offers several desirable features for implementing stabilised weights and generalised raking but also presents some challenges. In two-phase sampling, the relationships between design variables and sampling weights are explicitly known. Stratified sampling of individuals is typically used, rather than multistage cluster sampling. As a result, the sampling is likely to be strongly informative and, in particular, to depend on the outcome variable in the model of interest (Equation (\ref{the-model})). Consequently, simply ignoring the sampling is not satisfactory, and some design variables may be inappropriate to include in the model. Additionally, sampling designs may depend on additional variables $A$, such as $X$ measured with error. This raises the challenge when stabilisation of weights has to coexist with raking of weights, requiring a careful integration of both methods. For example, the additional variables $A$ can be incorporated into the generalised raking procedure but cannot be used when estimating the stabilised weights because they are not included in the model (Equation (\ref{the-model})).

The contribution of this paper is two-fold. First, we consider stabilised weights under case--control logistic regression, which provides an analytically tractable setting for studying efficiency, and show that full efficiency can be recovered in this special setting. Second, we extend the approach to general two-phase samples for regression models, showing how to combine stabilised weights, which multiply the design weights $d_i$ by a function of covariates $Z$, with raked weights, which scale $d_i$ by a function of the phase-1 variables $(Z,A,Y)$. We show that there are intrinsic barriers to combining them in full generality, but that combination is possible and beneficial for improving precision in some realistic settings. However, efficiency gains may be limited when the variation in design weights is small, as reflected by a small coefficient of variation and hence similar weights across sampled units. This is likely the case under the optimal designs studied by \citet{McIsaac2015Adaptive} and \citet{chen2020}.

Section~\ref{names} describes generalised raking and stabilised weights, reviews their properties, and derives the estimation procedure for combining the two modifications together with the corresponding standard error estimator. Section~\ref{simulations} evaluates the performance of these modifications through extensive simulation studies. Section~\ref{appl} applies the developed methods to a large multinational two-phase study of Kaposi sarcoma (KS) among people living with HIV.

In contrast to generalised raking, the stabilised weight estimator has not been shown to achieve full semiparametric efficiency under its assumptions.  Semiparametric maximum likelihood estimators, which are available for certain special cases of this two-phase sampling problem, can be more efficient \citep{Scott1997, Tao2017Efficient}, but they make stronger assumptions. We discuss the trade-offs associated with the use of more sophisticated estimators in Section~\ref{discussion}. The code for simulation studies is available at \url{https://www.github.com/t0ngchen/stabrake}.

\section{Weight modifications}
\label{names}

In this section, we provide an overview of generalised raking and stabilised weights, reviewing their properties and illustrating their use in estimating regression parameters. We then propose an estimation procedure that combines these two methods and derive the corresponding standard error estimator.

\subsection{Generalised raking}
\label{subsec-gr}
Generalised raking adjusts design weights $d_i$ using phase-1 auxiliary variables $G$ while ensuring that the weighted total of $G$ in the sample matches the known population total. Specifically, suppose we are interested in estimating the population total of variable $X$, the generalised raking estimator can be defined as $T_{x,gr} = \sum_{i=1}^{N} R_i d_i g(G_i;\alpha)  X_i$, where $g(G_i;\alpha)$ is a raking factor, defined as a function of the auxiliary variables with parameter $\alpha$, and $d_i g(G_i;\alpha)$ are the raked weights. The goal is to adjust the design weights to minimise total weight changes 
 $$
 \sum_{i=1}^{N} R_i \delta (d_i g(G_i;\hat{\alpha}), d_i ),
 $$
under a distance function $\delta(a,b)$, while ensuring the calibration constraints 
\begin{equation}
\sum_{i=1}^N R_i d_i g(G_i;\hat{\alpha}) G_i=\sum_{i=1}^N G_i.\label{calibration-constraints}
\end{equation}

The minimisation is typically done using Lagrange multipliers. The generalised regression estimator (GREG) can be obtained using distance function $\delta(a,b) = (a-b)^2/2b$. Importantly, generalised raking estimators are closely connected to augmented inverse-probability weighted (AIPW) estimators \citep{robins1994estimation}, which are developed from regression estimators of the population total \citep{deville1992calibration, robins1994estimation, lumley2011connections}. Like the IPW estimator, generalised raking estimators are a class of design-based estimators and therefore do not rely on assumptions about the outcome model. Additionally, generalised raking estimators are asymptotically efficient among design-based estimators \citep{deville1992calibration, lumley2011connections} and gain precision by leveraging auxiliary information. This efficiency gain depends on the linear correlation between the auxiliary variables and the variable for which the population total is being estimated. 

Our objective is to improve the efficiency of regression parameter estimation in regression models. To achieve this, it is useful to express the regression parameter $\beta$ as a total. An asymptotically linear estimator $\hat{\beta}$ can be written as 
\begin{equation}
    \sqrt{N}(\hat{\beta} - \beta) = \frac{1}{\sqrt{N}} \sum_{i=1}^{N} h_i(\beta) + o_p(1), \label{eif}
\end{equation}
where $h_i(\beta)$ denotes the influence function for the $i$th observation and its dependence on additional nuisance parameters is suppressed for notational simplicity. This formulation implies that $\hat{\beta}$ minus a constant can be viewed as an average of its influence functions. To maximise the efficiency gain for the estimates of regression parameters, auxiliary variables should then be strongly correlated with influence functions \citep{breslow2009using, Chen2022optimal}. Specifically, a generalised raking estimator is asymptotically efficient when the auxiliary variables are $G_i = E(h_i(\beta)\mid Y_i,Z_i,A_i)$, the conditional expectation of influence functions given phase-1 data \citep{breslow2009using, lumley2011connections}. Since influence functions are typically unknown, they need to be estimated. A common approach is the `plug-in' method \citep{Kulich2004improving}, which first imputes $X$ and then uses the imputed $X$ together with phase-1 data to estimate influence functions. This procedure may be improved by multiple imputation. However, similar performance has been observed for generalised raking when influence functions, used as auxiliary variables, are obtained through either single or multiple imputation \citep{Han2020Combining, Shepherd2023Multiwave}. The use of generalised raking for regression models has been well-established in the context of two-phase medical studies \citep{SAMUELSEN2007, breslow2009using, Ganna2012, breslow2013using, Noma2017Analysis, Shepherd2023Multiwave}.

\subsection{Stabilised weights}
Stabilised weights reduce non-essential variation in sampling weights by adjusting them to account for the variation explained by covariates. Specifically, stabilised weights are derived based on the principle that weights in a regression model can be multiplied by an arbitrary function of the covariates without introducing bias. However, this relies on an additional assumption that the marginal mean model in Equation~(\ref{the-model}) is correctly specified, so that for each observation $i$, 
\begin{equation*}\phi(E[Y_i|X_i,Z_i])=\beta_x X_i+\beta_z Z_i.\end{equation*}

Let $q(x,z)$ denote a non-negative arbitrary fixed function defined on the support of $(X,Z)$; when it depends only on $Z$, we write $q(z)$. Let $q(X_i,Z_i)$ denote the evaluation of $q(x,z)$ at the $i$th observation. Stabilising the design weights modifies the weight function to be $w = dq(x,z)$ and changes the population target inference from the solution $\beta_0$ of 
$$ 
E\left[U(Y_i,X_i,Z_i;\beta)\right]=0
$$
to the solution $\beta_q$ of
$$
E\left[d_i q(X_i,Z_i) U(Y_i,X_i,Z_i;\beta) \right]=0,
$$
where $U(Y_i,X_i,Z_i;\beta)$ denotes the estimating equation.  If we assume that $E[U(Y_i,X_i,Z_i;\beta)]=0$ for every unit in the population, which for generalised linear models is equivalent to correct specification of the marginal mean model, then we have $\beta_0=\beta_q$.

When stabilisation of weights is valid, both estimators $\hat{\beta}$ and $\hat{\beta_q}$ are consistent estimators of the same parameter $\beta_0$ in the underlying data-generating mechanism. However, this does not imply that $\hat{\beta}=\hat{\beta_q}$. The central benefit of using stabilised weights is to reduce unnecessary variability in the weights, leading to a more precise estimate.

If we replace the design weights with a modified weight function $w = dq(x,z)$ that is less variable as a function of $X$ and $Z$, the variances of the regression parameter estimates $(\hat\beta_x,\hat\beta_z)^\top$ will change. Our aim is to pick the function $q()$ to reduce the variances, i.e., to minimise $\text{var}(\hat{\beta}_q)$. Heuristically, in the case of a linear model, if the design weights could be written as $d_1(x,z)\times d_2(y)$, we would ideally choose $q(x,z)=1/d_1(x,z)$ to remove the ignorable component of variation from the weights. However, design weights will, in general, not be separable in this way, so more considerations are needed to estimate the function $q()$.

Under the assumption that observations for different units are independent, \cite{SKINNER2012} showed the variance $\text{var}(\hat{\beta}_q)$ can be written as   

\begin{equation*}
    \text{var}(\hat{\beta}_q) \approx  \left\{ J(\beta)\right\}^{-1} \sum_{i=1}^N q(X_i,Z_i)^2 E\left(d_i U_i(\beta) U_i(\beta)^T \right) \left\{J(\beta) \right\}^{-1},
\end{equation*}
where
\begin{equation*}
J(\beta) =  \sum_{i=1}^{N}  q(X_i,Z_i) E\left(\frac{\partial U_i}{\partial \beta}\right) 
\end{equation*}

For generalised linear models, we have $U(Y_i,X_i,Z_i;\beta) = {e_i(Y_i;X_i,Z_i,\beta) \cdot (X_i, Z_i)}$, where $e_i()$ is a scalar function. Under the canonical link, we have $e_i(Y_i;X_i,Z_i,\beta) = Y_i - E(Y_i \mid X_i, Z_i)$. For notational simplicity, we write $e_i = e_i(Y_i;X_i,Z_i,\beta)$.

\begin{equation*}
    \text{var}(\hat{\beta}_q) \approx  \left\{ J(\beta)\right\}^{-1} \sum_{i=1}^N q(X_i,Z_i)^2 E\left(d_i e_i^2\right) (X_i,Z_i) (X_i, Z_i)^T\left\{J(\beta) \right\}^{-1},
\end{equation*}
where
\begin{equation*}
J(\beta) = \sum_{i=1}^N q(X_i,Z_i) E\left(e_i^2\right) {(X_i,Z_i)} {(X_i,Z_i)}^T.
\end{equation*}
The optimal solution of the function $q()$ is then \citep{SKINNER2012}
\begin{equation}
    q(X_i,Z_i) \propto \frac{E\left(e_i^2 \mid X_i, Z_i\right)}{E\left(d_i e_i^2 \mid X_i, Z_i\right)}.\label{opt-stab}
\end{equation}

 The numerator and denominator of $q(X_i, Z_i)$ can be estimated by design-weighted auxiliary regressions of $\hat e_i^2$ and $d_i \hat e_i^2$ on $X_i$ and $Z_i$ respectively \citep{SKINNER2012}. \citet{SKINNER2012} derived the function $q()$ for generalised linear models, while the heteroscedastic form of linear regression was given by \citet{fuller2009sampling}. The results of \citet[Section 6.3.2]{fuller2009sampling} can be generalised to heteroscedastic generalised linear models, in which $q()$ takes a form similar to Equation~(\ref{opt-stab}), but the numerator and denominator are based on the model variance and the empirical variance, respectively.  Equation~(\ref{opt-stab}) thus represents a special case of this more general form.

\subsection{A two-step approach: Stabilise before raking} \label{two-step-sr}

In this section, we propose combining generalised raking with stabilised weights by choosing the function $q()$ to depend only on $Z$, rather than phase-2 variables $X$. This is because to satisfy the constraints in Equation~(\ref{stab-calibration-constraints}) below, the values of $q(X_i, Z_i)$ must be available for all individuals in the phase-1 sample. Since $X$ is only available in phase 2, the optimal choice of $q()$ that depends on $X$ is not clear. We therefore confine the function $q()$ to the phase-1 variable $Z$. Of note, stabilised weights can only be computed using variables included in the outcome model. Although generalised raking allows the use of imputed or error-prone phase-1 versions of $X$ \citep{breslow2009using, Amorim2021twophase}, these variables cannot be used to calculate the stabilised weights as they are not included in the outcome model.

The proposed method combines stabilised weights and generalised raking through a two-step procedure, where we stabilise the design weights before applying generalised raking. Specifically, we first compute the stabilisation term $q(Z_i)$ according to Equation~(\ref{opt-stab}) using phase-1 variables in the outcome model, so we have 
\begin{equation}
    q(Z_i) \propto \frac{E\left(e_i^2 \mid Z_i\right)}{E\left(d_i e_i^2 \mid Z_i\right)}.\label{opt-stab-z}
\end{equation}
The data can then be represented as a phase-1 sample with weights $q(Z_i)$ and a phase-2 sample with the actual design weights $d_i = 1/\pi_i$.

Subsequently, the second step is to apply generalised raking based on auxiliary variables $G$. The stabilised calibration constraints are
\begin{equation}
\sum_{i=1}^N R_id_ig(G_i;\alpha)q(Z_i)G_i=\sum_{i=1}^N q(Z_i)G_i,\label{stab-calibration-constraints}
\end{equation}
where $q(Z_i)$ appears on both sides of the equation. Generalised raking, as usual, will lead to a variance that is asymptotically no larger than that obtained with the unraked weights, and smaller to the extent that $q(z)G$ is correlated with influence functions. The steps of the procedure are outlined as follows:

\begin{enumerate}
    \item  Estimate the stabilisation term $q(Z_i)$, which depends only on the covariates $Z$ rather than $X$, based on Equation~(\ref{opt-stab-z}). The values of $q(Z_i)$ are calculated for all individuals in the phase-1 sample.  
    
    \item Construct the auxiliary variables $G$ as described in Section \ref{subsec-gr}. Since our goal is the estimation of regression parameters, the auxiliary variables $G$ should ideally be the influence functions of regression parameters. These influence functions can be estimated following the steps described in \cite{Kulich2004improving} and \cite{breslow2009using}.
    
    \item Calculate $g(G;\alpha)$ that satisfies the calibration constraints in Equation~(\ref{stab-calibration-constraints}) by solving the same Lagrange multiplier problem as in the standard generalised raking procedures described in Section~\ref{subsec-gr}. 

    \item Fit the weighted regression model with raked weights calculated from Step 3.

\end{enumerate}
The proposed approach can be implemented using standard statistical software that supports two-phase samples and generalised raking. It can be implemented using the \texttt{survey} package \citep{Lumley2010complex} in \texttt{R} \citep{Rcore} by specifying a two-phase design using the \texttt{twophase} function. In Step 3, we use a computation trick to get the desired adjustments in the \texttt{survey} package. Specifically, we set the phase-1 probability to $\epsilon/q(z)$ where $\epsilon$ is a sufficiently small constant, and we set the phase-2 probability as $\pi$. We can then calculate $g(G;\alpha)$ by applying the \texttt{calibrate} function to this two-phase design.
Detailed examples of the implementation are available at \url{https://github.com/t0ngchen/stabrake}. In the following section, we derive the variance estimator for the proposed approach.
 
\subsection{Variance estimation}
In this section, we derive the variance estimator for the proposed two-step approach. Here, auxiliary variables $G$ are influence functions constructed from $(Y,Z,A)$. Accordingly, we write the raking factor as $g(Y,Z,A;\alpha)$.

The generalised raking and weights stabilisation adjustments depend on phase-1 data and on parameters $\alpha$ in $g(Y,Z,A;\alpha)$ and $\gamma$ in $q(x;\gamma)$, respectively. For every fixed value of $\alpha$ and $\gamma$ the weighted estimating equation
\begin{equation}
\sum_{i=1}^N R_i d_i q(Z_i;\gamma) g(Y_i,Z_i,A_i;\alpha) U(Y_i,X_i,Z_i;\beta)=0
\end{equation}
is unbiased for the model parameter $\beta_0$ if the finite population is generated by the outcome model.

\paragraph{Proposition}
Suppose there exist limiting values $\alpha^*$ and $\gamma^*$ such that $\|\hat\alpha-\alpha^*\|_2=o_p(n^{-1/4})$ and $\|\hat\gamma-\gamma^*\|_2=o_p(n^{-1/4})$. By standard `plug-in' arguments \citep{Newey1994Econ}, the solution to 
\begin{equation}
\sum_{i=1}^N R_i d_i q(Z_i;\hat\gamma) g(Y_i,Z_i,A_i;\hat\alpha) U(Y_i,X_i,Z_i;\beta)=0
\end{equation}
is consistent for $\beta_0$ and asymptotically Normal, with asymptotic variance depending on $\hat\alpha$ and $\hat\gamma$ only through $\alpha^*$ and $\gamma^*$. That is,
\begin{equation}
\sqrt{n}(\hat\beta-\beta_0)\stackrel{d}{\to}N(0, V(\alpha^*,\gamma^*)).
\end{equation}
\qed

\paragraph{Computational formula} When the first phase is sampled from a large population or a data-generating process, $V(\alpha,\gamma)$ is the same as the asymptotic design variance of $\hat\beta$ if the data had come from a two-phase design where phase-1 observations were independently sampled with probability proportional to $1/q(z;\gamma)$ and phase 2 was sampled with its actual design and then raked on the auxiliary variables to get adjustments $g(Y_i,Z_i,A_i;\alpha)$. For notational simplicity, we write $q_i = q(Z_i;\gamma)$.

That is, let $\tilde q^{-1}_i=\epsilon q^{-1}_i$ where $\epsilon>0$ is chosen small enough that $\tilde q^{-1}_i(1-\tilde q^{-1}_i)$ can be approximated well by $\tilde q^{-1}_i$. The exact value will turn out not to matter. 

Let $\Delta_{ij} = \mathrm{Cov}(R_i, R_j)$ \citep{sarndal-book}.
The phase-1 covariance of sampling indicators is 
$\Delta_{1ij}=0$ if $i\neq j$ and $\Delta_{1ii}=\tilde q^{-1}_i(1-\tilde q^{-1}_i)$. Define $\pi^*_{ij}=\tilde q^{-1}_i\tilde q^{-1}_j\pi_{ij}$. 

Following \citet[Result 9.7.1]{sarndal-book}, 
the phase-1 contribution to the variance of the super-population total of $U_i(\beta)$ is estimated by
$$\sum_{i=1}^N R_i \frac{\Delta_{1ii}}{\pi^*_{ii}} \tilde q_i U_i(\hat\beta)\tilde q_i U_i(\hat\beta)^T,$$
which we can approximate by
$$\sum_{i=1}^N R_i\frac{\tilde q_i^2 }{\pi_{ii}} U_i(\hat\beta)U_i(\hat\beta^T).$$

The phase-2 contribution to the variance is 
$$\sum_{i,j=1}^N R_{ij} \frac{\Delta_{ij}}{\pi_{ij}}g_{i}(\alpha)g_j(\alpha)d_id_j\tilde q_i(\gamma)\tilde q_j(\gamma)U_i(\hat\beta)U_j(\hat\beta)$$
where $\pi_{ij}$ is the probability that both $i$ and $j$ were subsampled given the observed phase-1 sample and $\Delta_{ij}=\pi_{ij}-\pi_i\pi_j$ is the sampling indicator covariance.

The mean of $U(\beta)$ is the total divided by $N/\epsilon$, so we may cancel a factor of $\epsilon^{-2}$ from the numerator and denominator. The asymptotic variance of the mean of $U(\beta)$ can thus be estimated  by 

$$\hat V(\alpha,\gamma) = \frac{\epsilon^2}{N^2}\sum_{i=1}^N R_i\frac{ \tilde q_i^2 }{\pi^*_{ii}} U_i(\hat\beta)U_i(\hat\beta^T)+\sum_{i,j=1}^N R_{ij} \frac{\Delta_{ij}}{\pi_{ij}}g_{i}(\alpha)g_j(\alpha)d_id_j \tilde q_i(\gamma) \tilde q_j(\gamma)U_i(\hat\beta)U_j(\hat\beta)$$
 or
$$\hat V(\alpha,\gamma) = \frac{1}{N^2}\sum_{i=1}^N R_i\frac{ q_i^2 }{\pi_{ii}} U_i(\hat\beta)U_i(\hat\beta^T)+\sum_{i,j=1}^N R_{ij} \frac{\Delta_{ij}}{\pi_{ij}}g_{i}(\alpha)g_j(\alpha)d_id_j q_i(\gamma) q_j(\gamma)U_i(\hat\beta)U_j(\hat\beta)$$
 
\qed

This result allows computation of $V$ with standard software that handles two-phase samples and generalised raking.

\section{Simulations}
\label{simulations}

We present two groups of simulations to evaluate the performance of stabilised weights \citep{SKINNER2012}, as well as our proposed two-step approach that combines stabilised weights with generalised raking in the context of two-phase samples. Specifically, we evaluate two versions of stabilised weights: one with the stabilisation term $q(z)$ that stabilises only on $Z$ based on Equation~(\ref{opt-stab-z}), and another with $q(x,z)$ that stabilises on both $X$ and $Z$ based on Equation~(\ref{opt-stab}). Note that $q(x,z)$ can only be calculated for individuals in the phase-2 sample because $X$ is not available at phase 1. We also assess the performance of our proposed two-step approach, in which weight stabilisation is applied only to $Z$, and the weights are calibrated using influence functions as auxiliary variables to obtain generalised raking estimators.

First, we consider logistic regression under case--control sampling, which can be viewed as a two-phase design in which individuals are sampled conditional on the outcome. The second group of simulations considers general stratified two-phase designs, including a balanced design with equal allocation across strata \citep{Breslow1988} and an optimal design for design-based estimators \citep{Chen2022optimal}. We consider continuous outcomes with homoscedastic and heteroscedastic errors. A setting with a binary outcome is also considered. Across all simulation scenarios, the bias was negligible with $\lvert \text{bias} \rvert < 0.01$. Therefore, performance was evaluated in terms of the empirical standard error (empSE) and the root mean squared error (RMSE).

\subsection{Case--control sampling}

Stabilised weights improve efficiency by removing non-essential variation in the weights. However, the extent to which efficiency gains can be achieved is generally unclear. In most settings, this cannot be assessed analytically. For case--control sampling with logistic regression, however, the stabilised weights have an analytical form, and the efficient estimator is known \citep{Prentice1979Logistic}. This enables a direct assessment of efficiency analytically.

\subsubsection{Stabilised weights under case--control sampling} \label{cc-closed-form}
Following Equation~(\ref{opt-stab}) we derive the optimal form of the function $q()$ for the case--control sampling. Let $p(X,Z) = \Pr(Y=1 \mid X,Z)$ denote the predicted probability. For logistic regression in case--control sampling, the numerator in Equation~(\ref{opt-stab}) is $E\left[(Y-p(X,Z))^2 \mid X,Z\right] = p(X,Z) (1-p(X,Z))$, and the denominator can be calculated as:
\begin{align*}
E\left[\pi^{-1}(Y-p(X,Z))^2 \mid X,Z\right] &= E \left[ Y(Y-p(X,Z))^2 + \frac{N - n_{\text{cases}}}{n_{\text{cases}}} \left(1-Y\right) \left(Y-p(X,Z)\right)^2 \mid X,Z \right]\\
& = p(X,Z)\left(1-p(X,Z)\right)^2 + \frac{N - n_{\text{cases}}}{n_{\text{cases}}} p(X,Z)^2 (1-p(X,Z)).
\end{align*}
The optimal function $q()$ for the case-control sampling can then be written explicitly as
\begin{equation}
    q(X_i,Z_i) = \frac{1}{(1 - p(X_i,Z_i)) + \left((N - n_{\text{cases}})/n_{\text{cases}}\right) p(X_i,Z_i)}.\label{cc-sm}
\end{equation}
Based on Equation~(\ref{cc-sm}), in the scenario where stabilisation is applied to $Z$ only,
$q(Z_i)$ is obtained by simplifying the above equation as
\begin{equation}
    q(Z_i) = \frac{1}{(1 - p(Z_i)) + \left((N - n_{\text{cases}})/n_{\text{cases}}\right) p(Z_i)}.\label{cc-sm1}
\end{equation}

\subsubsection{Simulation design}

We simulated a population of $N = 10{,}000$ individuals and drew 1--to--1 case--control samples. We focused on the special setting with a single exposure. Specifically, the exposure $X$ was generated from either a normal or a uniform distribution. We compared IPW and stabilised weight (according to Equation~\eqref{cc-sm}) estimators with the maximum likelihood estimator (MLE), where the intercept was corrected using the known sampling fractions for cases and controls \citep{Anderson1972, Prentice1979Logistic}.
 
We first considered a rare disease scenario. The continuous exposure $X$ was drawn from a standard normal distribution, $X \sim N(0,1)$. The binary outcome $Y$ was simulated using the logistic regression model $\text{logit}(P(Y = 1 \mid X)) = \beta_0 + \beta_x X$, where $\text{logit}(x) = \text{log}(x/(1-x))$ and the intercept was set to $\beta_0 = -4$. The coefficient $\beta_x$ was varied to evaluate the performance of the proposed estimator across different effect sizes of exposure $X$. $1000$ case--control samples were simulated.

\begin{table}[h]
\centering
\begin{tblr}
{
colspec={Q[]Q[]Q[]Q[]Q[]Q[]Q[]Q[]Q[]},
cell{1}{4}={c=2,}{halign=c,},
cell{1}{6}={c=2,}{halign=c,},
cell{1}{8}={c=2,}{halign=c,},
cell{3}{1}={r=2,}{valign=h,},
cell{5}{1}={r=2,}{valign=h,},
cell{7}{1}={r=2,}{valign=h,},
cell{9}{1}={r=2,}{valign=h,},
cell{3}{2}={r=2,}{valign=h,},
cell{5}{2}={r=2,}{valign=h,},
cell{7}{2}={r=2,}{valign=h,},
cell{9}{2}={r=2,}{valign=h,},
}
\toprule
& &  & MLE &  & Stab &  & IPW &  \\
\cmidrule[lr]{4-5}\cmidrule[lr]{6-7}\cmidrule[lr]{8-9}
$\Bar{n}$ & $\beta_x$ &  & empSE & RMSE & empSE & RMSE & empSE & RMSE \\
\midrule
360 & 0 & Intercept & 0.076 & 0.076 & 0.076 & 0.076 & 0.076 & 0.076 \\
360 & 0 & X         & 0.103 & 0.103 & 0.103 & 0.102 & 0.104 & 0.104 \\
403 & 0.5 & Intercept & 0.082 & 0.082 & 0.082 & 0.082 & 0.083 & 0.084 \\
403 & 0.5 & X         & 0.110 & 0.110 & 0.110 & 0.110 & 0.119 & 0.119 \\
560 & 1 & Intercept & 0.089 & 0.090 & 0.089 & 0.090 & 0.101 & 0.102 \\
560 & 1 & X         & 0.108 & 0.108 & 0.107 & 0.107 & 0.138 & 0.139 \\
878 & 1.5 & Intercept & 0.096 & 0.096 & 0.096 & 0.096 & 0.118 & 0.119 \\
878 & 1.5 & X         & 0.107 & 0.107 & 0.107 & 0.107 & 0.140 & 0.142 \\
\bottomrule
\end{tblr}
\caption{Empirical standard error (empSE) and root mean squared error (RMSE) for different estimators under varying values of $\beta_x$. $\Bar{n}$ represents the average phase-2 sample size over all simulation replicates. Three estimators are compared: MLE, IPW, and the stabilised weight estimator (Stab).}
\label{cc-norm}
\end{table}

The results are presented in Table~\ref{cc-norm}. When $\beta_x = 0$, all three estimators achieved the same efficiency. As $\beta_x$ increased, the stabilised weight estimator (Stab) maintained the same efficiency as the MLE. They were more efficient than the IPW estimator, particularly at larger $\beta_x$ values. In Supplementary Material Section S1, we considered a similar case-control design with uniformly distributed exposure $X$, where the IPW estimator would be expected to perform better because $X$ was evenly spread. Table S1 showed a similar pattern to the previous simulations, except that the efficiency of the IPW estimator was close to that of the other three estimators.

For case--control sampling without phase-1 information \citep{Prentice1979Logistic}, we showed that stabilised weights can recover full efficiency under correct model specification. More generally, the analytical form of the optimal function $q()$ was unknown for general stratified two-phase designs, including case--control sampling with phase-1 covariates, and therefore had to be estimated. In Supplementary Material Section S2, we considered this setting and showed that the proposed two-step approach and stabilised weights improved efficiency upon generalised raking and IPW, respectively.

\subsection{General two-phase designs}\label{sim-gen-tp}

In this section, we evaluate the performance of the proposed methods under general stratified two-phase designs. Specifically, we considered the IPW estimator, generalised raking (GR) calibrated on influence functions of the outcome model using the `plug-in' method described in Section~2.1, the proposed two-step approach that incorporated stabilised weights into generalised raking as described in Section~\ref{two-step-sr}, and two stabilised weight estimators (Stab$_{xz}$ and Stab$_{z}$). The Stab$_{xz}$ estimator used $q(x,z)$ for stabilization, as defined in Equation~(\ref{opt-stab}), whereas Stab$_{z}$ relied on $q(z)$, as defined in Equation~(\ref{opt-stab-z}). Unlike the special case--control designs, closed-form expressions for $q()$ are generally unavailable under general two-phase designs. As suggested by \citet{SKINNER2012}, we estimated the numerator and denominator of $q()$ via design-weighted auxiliary regressions, using $\hat e_i^{\,2}$ and $d_i \hat e_i^{\,2}$ as response variables respectively. Specifically, we estimated $q(x,z)$ using a generalised additive model \citep{hastie1986gam}, and $q(z)$ using Friedman’s SuperSmoother \citep{friedman1984techreport}. We began with continuous outcomes with homoscedastic errors and extended to scenarios with heteroscedastic errors. Extensive simulation studies more informative designs were presented in Supplementary Material Sections S3–S4.

\subsubsection{Continuous outcome with homoscedastic errors}

We considered a stratified two-phase sampling design with a continuous outcome, where stratification was based on covariate $Z$. 1000 datasets with $N = 6000$ individuals were generated. The variables were generated as follows: $W_1 \sim N(0,1)$ and $W_2 \sim \text{Bernoulli}(0.6)$. The phase-2 variable $X$ was defined as $X = 0.3 + W_1 + W_2 + \epsilon_X$, with $\epsilon_X \sim N(0,1)$, and the phase-1 covariate $Z$ was defined as $Z = W_1 + 0.8 \times W_2 + \epsilon_Z$, with $\epsilon_Z \sim N(0,1)$. An additional variable $A$ was defined as $A = X + \epsilon_A$, with $\epsilon_A \sim N(0,1)$ and was used in the generalised raking. The outcome was generated from the linear model $Y = \beta_x X + \beta_z Z + \epsilon_Y$, where $\epsilon_Y \sim N(0,\sigma^2)$. We set $\beta_x = \beta_z$ and varied their values, as well as the residual variance $\sigma^2$. Stratification was based on the 90th percentile of $Z$. A balanced phase-2 subsampling design was applied, and 600 individuals were selected for the phase-2 sample.

\begin{table}[h]
\centering
\resizebox{\textwidth}{!}{%
\begin{tblr}
{
colspec={Q[]Q[]Q[]Q[]Q[]Q[]Q[]Q[]Q[]Q[]Q[]Q[]},
cell{1}{4}={c=2,}{halign=c,},
cell{1}{6}={c=2,}{halign=c,},
cell{1}{8}={c=2,}{halign=c,},
cell{1}{10}={c=2,}{halign=c,},
cell{1}{12}={c=2,}{halign=c,},
cell{3}{2}={r=2,}{valign=h,},
cell{5}{2}={r=2,}{valign=h,},
cell{7}{2}={r=2,}{valign=h,},
cell{9}{2}={r=2,}{valign=h,},
cell{11}{2}={r=2,}{valign=h,},
cell{13}{2}={r=2,}{valign=h,},
cell{15}{2}={r=2,}{valign=h,},
cell{17}{2}={r=2,}{valign=h,},
cell{3}{1}={r=8,}{valign=h,},
cell{11}{1}={r=8,}{valign=h,},
}
\toprule
&  &  & IPW &  & Stab$_{xz}$ &  & Stab$_z$ &  & Stab-rake &  & GR &  \\
\cmidrule[lr]{4-5}\cmidrule[lr]{6-7}\cmidrule[lr]{8-9}\cmidrule[lr]{10-11}\cmidrule[lr]{12-13}
$\sigma^2$ & $\beta_x \& \beta_z$ &  & empSE & RMSE & empSE & RMSE & empSE & RMSE & empSE & RMSE & empSE & RMSE \\
\midrule
0.5 & 0.1 & X & 0.029 & 0.029 & 0.027 & 0.027 & 0.024 & 0.024 & 0.019 & 0.019 & 0.019 & 0.019 \\
0.5 & 0.1 & Z & 0.026 & 0.026 & 0.023 & 0.023 & 0.021 & 0.021 & 0.016 & 0.016 & 0.012 & 0.012 \\
0.5 & 0.5 & X & 0.029 & 0.029 & 0.027 & 0.027 & 0.024 & 0.024 & 0.020 & 0.020 & 0.022 & 0.022 \\
0.5 & 0.5 & Z & 0.026 & 0.026 & 0.023 & 0.023 & 0.021 & 0.021 & 0.018 & 0.018 & 0.018 & 0.018 \\
0.5 & 1.0 & X & 0.029 & 0.029 & 0.027 & 0.027 & 0.024 & 0.024 & 0.022 & 0.022 & 0.025 & 0.025 \\
0.5 & 1.0 & Z & 0.026 & 0.026 & 0.023 & 0.023 & 0.021 & 0.021 & 0.019 & 0.019 & 0.022 & 0.022 \\
0.5 & 1.5 & X & 0.029 & 0.029 & 0.027 & 0.027 & 0.024 & 0.024 & 0.023 & 0.023 & 0.027 & 0.027 \\
0.5 & 1.5 & Z & 0.026 & 0.026 & 0.023 & 0.023 & 0.021 & 0.021 & 0.020 & 0.020 & 0.024 & 0.024 \\
1.0 & 0.1 & X & 0.041 & 0.041 & 0.037 & 0.037 & 0.033 & 0.033 & 0.029 & 0.029 & 0.027 & 0.027 \\
1.0 & 0.1 & Z & 0.037 & 0.037 & 0.032 & 0.032 & 0.029 & 0.029 & 0.027 & 0.027 & 0.017 & 0.017 \\
1.0 & 0.5 & X & 0.041 & 0.041 & 0.037 & 0.037 & 0.033 & 0.033 & 0.030 & 0.030 & 0.029 & 0.029 \\
1.0 & 0.5 & Z & 0.037 & 0.037 & 0.032 & 0.032 & 0.029 & 0.029 & 0.028 & 0.028 & 0.022 & 0.022 \\
1.0 & 1.0 & X & 0.041 & 0.041 & 0.037 & 0.037 & 0.033 & 0.033 & 0.031 & 0.031 & 0.033 & 0.033 \\
1.0 & 1.0 & Z & 0.037 & 0.037 & 0.032 & 0.032 & 0.029 & 0.029 & 0.028 & 0.028 & 0.028 & 0.028 \\
1.0 & 1.5 & X & 0.041 & 0.041 & 0.037 & 0.037 & 0.033 & 0.033 & 0.032 & 0.032 & 0.036 & 0.036 \\
1.0 & 1.5 & Z & 0.037 & 0.037 & 0.032 & 0.032 & 0.029 & 0.029 & 0.028 & 0.028 & 0.032 & 0.032 \\
\bottomrule
\end{tblr}}
\caption{Empirical standard error (empSE) and root mean squared error (RMSE) of $X$ and $Z$ for different estimators under varying values of $\beta_x$, $\beta_z$ and $\sigma^2$. Five estimators are compared: IPW, generalised raking (GR), the proposed two-step approach that incorporates stabilised weights into generalised raking (Stab-rake), and two stabilised weight estimators (Stab$_{xz}$ and Stab$_{z}$). The Stab$_{xz}$ estimator uses $q(x,z)$ as defined in Equation~(\ref{opt-stab}), while the Stab$_{z}$ estimator uses $q(z)$ as defined in Equation~(\ref{opt-stab-z}).}
\label{sim3}
\end{table}

The results are presented in Table~\ref{sim3}. The stabilised weight estimators (Stab$_{xz}$ and Stab$_{z}$) were more efficient than the IPW estimator but less efficient than the proposed two-step estimator (Stab-rake). Stab$_{xz}$ was slightly less efficient than Stab$_{z}$, which might be due to the relatively poorer performance of the two-dimensional smoothing used to estimate $q(x,z)$. These results illustrate how stabilisation becomes more difficult as the dimension of the conditioning variables increases. When the smooth function in the denominator of equation~(\ref{opt-stab}) is close to zero, underestimation leads to inflation of the ratio, and this is challenging to control. \citet{SKINNER2012} avoided this issue by stabilising their cross-national samples only on country variables. For estimation of $\beta_x$, the proposed two-step approach (Stab-rake) was more efficient than generalised raking, and efficiency gain was greater for larger values of $\beta_x$ and $\beta_z$. For estimation of $\beta_z$, the generalised raking estimator was more efficient at smaller values of $\beta_x$ and $\beta_z$, while the two-step approach achieved higher efficiency when these parameters were large.

\subsubsection{Continuous outcome with heteroscedastic errors}

Building on the previous setting, we allowed the error variance to depend on covariates to assess performance under heteroscedasticity. Similarly, 1000 datasets with $N = 6000$ individuals were generated. The variables were generated as follows: $W_1 \sim N(0,1)$ and $W_2 \sim \text{Bernoulli}(0.6)$. The phase-2 variable $X$ was defined as $X = 0.3 + W_1 + W_2 + \epsilon_X$, with $\epsilon_X \sim N(0,1)$, and the phase-1 covariate $Z$ was defined as $Z = W_1 + 0.8 \times W_2 + \epsilon_Z$, with $\epsilon_Z \sim N(0,1)$. An additional variable $A$ was defined as $A = X + \epsilon_A$, with $\epsilon_A \sim N(0,1)$ and was used in the generalised raking. The outcome was generated from the linear model $Y = X + Z + \epsilon_Y$, where $\epsilon_Y \sim N(0, 1+\delta Z^2 )$. Stratification was based on the cross-classification of $A$ and $Z$, where $A$ was divided at its 25th and 75th percentiles and $Z$ at its 60th percentile, resulting in 6 strata in total. The optimal design for the design-based estimators with constant variance \citep{Chen2022optimal} was applied, and 600 individuals were selected for the phase-2 sample.

\begin{table}[htbp]
\centering
\resizebox{\textwidth}{!}{%
\begin{tblr}
{
colspec={Q[]Q[]Q[]Q[]Q[]Q[]Q[]Q[]Q[]Q[]Q[]Q[]},
cell{1}{3}={c=2,}{halign=c,},
cell{1}{5}={c=2,}{halign=c,},
cell{1}{7}={c=2,}{halign=c,},
cell{1}{9}={c=2,}{halign=c,},
cell{1}{11}={c=2,}{halign=c,},
cell{3}{1}={r=2,}{valign=h,},
cell{5}{1}={r=2,}{valign=h,},
cell{7}{1}={r=2,}{valign=h,},
cell{9}{1}={r=2,}{valign=h,},
cell{11}{1}={r=2,}{valign=h,},
cell{13}{1}={r=2,}{valign=h,},
}
\toprule
&  & IPW &  & Stab$_{xz}$ &  & Stab$_{z}$ &  & Stab-rake &  & GR &  \\
\cmidrule[lr]{3-4}\cmidrule[lr]{5-6}\cmidrule[lr]{7-8}\cmidrule[lr]{9-10}\cmidrule[lr]{11-12}
$\delta$ &  & empSE & RMSE & empSE & RMSE & empSE & RMSE & empSE & RMSE & empSE & RMSE \\
\midrule
0.10 & X & 0.036 & 0.036 & 0.035 & 0.035 & 0.035 & 0.035 & 0.029 & 0.029 & 0.029 & 0.029 \\
0.10 & Z & 0.044 & 0.044 & 0.043 & 0.043 & 0.043 & 0.043 & 0.031 & 0.031 & 0.030 & 0.030 \\
0.25 & X & 0.040 & 0.040 & 0.039 & 0.039 & 0.038 & 0.038 & 0.031 & 0.031 & 0.031 & 0.031 \\
0.25 & Z & 0.053 & 0.053 & 0.049 & 0.049 & 0.049 & 0.049 & 0.033 & 0.033 & 0.032 & 0.032 \\
0.50 & X & 0.047 & 0.047 & 0.042 & 0.042 & 0.041 & 0.041 & 0.034 & 0.034 & 0.036 & 0.036 \\
0.50 & Z & 0.067 & 0.067 & 0.057 & 0.057 & 0.057 & 0.057 & 0.038 & 0.038 & 0.037 & 0.037 \\
0.75 & X & 0.052 & 0.052 & 0.045 & 0.045 & 0.044 & 0.044 & 0.037 & 0.037 & 0.041 & 0.041 \\
0.75 & Z & 0.076 & 0.076 & 0.063 & 0.063 & 0.062 & 0.062 & 0.041 & 0.041 & 0.039 & 0.039 \\
1.00 & X & 0.058 & 0.058 & 0.048 & 0.048 & 0.046 & 0.046 & 0.040 & 0.040 & 0.046 & 0.046 \\
1.00 & Z & 0.087 & 0.087 & 0.069 & 0.069 & 0.069 & 0.069 & 0.046 & 0.046 & 0.042 & 0.042 \\
1.50 & X & 0.067 & 0.067 & 0.054 & 0.054 & 0.052 & 0.052 & 0.045 & 0.045 & 0.052 & 0.053 \\
1.50 & Z & 0.101 & 0.101 & 0.080 & 0.080 & 0.079 & 0.079 & 0.052 & 0.052 & 0.047 & 0.047 \\
\bottomrule
\end{tblr}}
\caption{Empirical standard error (empSE) and root mean squared error (RMSE) of $X$ and $Z$ for different estimators under varying values of $\delta$. Five estimators are compared: IPW, generalised raking (GR), the proposed two-step approach that incorporates stabilised weights into generalised raking (Stab-rake), and two stabilised weight estimators (Stab$_{xz}$ and Stab$_{z}$). The Stab$_{xz}$ estimator uses $q(x,z)$ as defined in Equation~(\ref{opt-stab}), while the Stab$_{z}$ estimator uses $q(z)$ as defined in Equation~(\ref{opt-stab-z}).}
\label{sim5}
\end{table}

The results are presented in Table~\ref{sim5}. When $\delta$ was small, the stabilised weight estimators were as efficient as the IPW estimator, and the proposed two-step estimator was as efficient as generalised raking. As $\delta$ increased, the efficiency gain from weight stabilisation became more substantial, with the stabilised weight estimators and the proposed two-step estimator achieving greater efficiency relative to IPW and generalised raking, respectively. 

In this section, we considered the optimal design derived under the assumption of constant error variance. However, since the model was correctly specified, those weights remained valid for the heteroscedastic estimand, because they are equivalent under this setting. 

\subsubsection{Additional simulation under more informative designs}

We conducted additional simulation studies under more informative two-phase designs, including optimal designs for design-based estimators \citep{Chen2022optimal}. Across these settings, efficiency gains from weight stabilisation were limited. This is consistent with reduced variability of the sampling weights under more informative designs, leaving less scope for efficiency improvement through weight stabilisation. Detailed results are provided in Supplementary Material Sections~S3--S4.

\section{Application to Kaposi’s sarcoma study}\label{appl}

We illustrate our methods using two-phase data from a study investigating predictors of KS at enrollment in care among people with HIV (PWH) in Latin America and East Africa from 2010-2019. Error-prone phase-1 data were available for $N = 177, 594$ PWH who met inclusion criteria. Because of data quality concerns, chart reviews were performed on a phase-2 sample of $n = 682$. The phase-2 sample was a stratified random sample. Strata were based on combinations of phase-1 measures of KS status and year of diagnosis (no KS, KS from 2010-2014, KS from 2015-2019) and study site (13 sites); some strata were collapsed due to small numbers resulting in a total of 34 strata. The number sampled from each stratum was based on an optimal design for design-based estimators, where the sampling design sought to maximise precision of the adjusted calendar year coefficient for KS incidence. More details about the cohorts and KS design are provided elsewhere \citep{amorim2025design}. 

Our goal for the adjusted KS prevalence analysis is to fit a logistic regression model with outcome $Y =$ KS status, covariates $X =$ sex, age, calendar year, prior use of antiretroviral therapy (ART), CD4 count (square-root transformed), and covariate $Z =$ region (Latin America vs. East Africa). $(Y, X)$ are only available in the phase-2 sample; error-prone versions of these variables, denoted $(Y^*,X^*)$ are available on everyone in the phase-1 cohort. Primary analyses for this study were based on generalised raking, calibrating the (estimated) design weights with the estimated influence function based on the error-prone data as auxiliary variables, $G_i=E(h_i(\beta)|Y_i^*,X_i^*,Z_i )$ \citep{slone2025ksanalysis}; accompanying IPW analyses were also performed. 

In analyses incorporating weight stabilisation, we stabilised on region, which compares the odds of KS at diagnosis between the Latin American (reference) and East African regions. Region, $Z$, is in the model of interest and error-free, so it can be used to stabilise the weights. The sampling design was strongly informative for KS diagnosis and region, but it was not focused on optimising precision for the prevalence-region odds ratio; thus, while large improvements in precision were not expected, we felt there was potential for modest gains in precision for estimates of the region coefficient. 

The general approach for weight stabilisation in studying KS follows from Sections \ref{two-step-sr} and \ref{sim-gen-tp}, in which we stabilised the weights prior to raking or IPW estimation. Having estimated $q(z)$ based on covariate $Z$, we fit the weighted regression with stabilised weights for IPW. For generalised raking, we used the estimated influence functions $G_i=E(h_i(\beta)|Y_i^*,X_i^*,Z_i )$ for all covariates from our outcome model of eligible phase-1 data \citep{Kulich2004improving}. Our stabilised calibration constraint includes $G_i$ and $q(Z_i)$ from phase 1 as in Equation (\ref{stab-calibration-constraints}), and $g(G_i;\alpha)$ are obtained by minimising the distance subject to the stabilised calibration constraint. We then used the raked weights in a weighted regression.

Results of our weighted approaches with and without weight stabilisation are presented in Table \ref{KS}. Generalised raking with stabilised weights (Stab-rake) narrowed the width of the 95\% CI of the log-odds ratio for region by 4.1\% over standard generalised raking. Generalised raking with stabilised weights narrowed the 95\% CI of the log-odds ratio for region by 19.0\% and 13.5\% over standard IPW and IPW with stabilised weights, respectively. Stabilisation had little impact on the precision of other covariate coefficients, aside from CD4 count and ART status. Lack of improvement was expected because these covariates could not be used to stabilise weights. CD4 count improved in precision to approximately the same degree as the estimate for Region. For the coefficient of ART status, stabilising on region reduced the width of the generalised raking 95\% CI by 6.6\%, which may be explained by differences in the weights across ART status levels.

\begin{table}[htbp]
\centering
\resizebox{\textwidth}{!}{%
\begin{tabular}{lcccccccc}
\toprule
& \multicolumn{2}{c}{IPW} 
& \multicolumn{2}{c}{Stab$_z$}
& \multicolumn{2}{c}{GR} 
& \multicolumn{2}{c}{Stab-rake} \\
\cmidrule(lr){2-3} \cmidrule(lr){4-5} \cmidrule(lr){6-7} \cmidrule(lr){8-9}
 & OR & 95\% CI & OR & 95\% CI & OR & 95\% CI & OR & 95\% CI \\
\midrule
Region East Africa (ref: Latin America)
 & 0.56 & 0.32--1.00 
 & 0.53 & 0.31--0.90
 & 0.73 & 0.45--1.19
 & 0.68 & 0.42--1.08 \\

Calendar Year
 & 0.96 & 0.86--1.06
 & 0.98 & 0.88--1.09
 & 0.86 & 0.78--0.94
 & 0.87 & 0.80--0.95 \\

Male sex (ref: female)
 & 2.27 & 1.30--3.97
 & 2.26 & 1.28--3.98
 & 2.65 & 1.99--3.54
 & 2.65 & 1.98--3.54 \\

Age (per 10 years)
 & 1.09 & 0.86--1.38
 & 1.08 & 0.85--1.38
 & 1.00 & 0.88--1.14
 & 1.01 & 0.89--1.15 \\

ART status (ref: no ART)
 & 2.84 & 1.05--7.64
 & 2.65 & 0.93--7.60
 & 4.15 & 2.30--7.46
 & 3.99 & 2.31--6.91 \\

CD4 count (sqrt-transformed)
 & 0.92 & 0.88--0.96
 & 0.92 & 0.88--0.96
 & 0.90 & 0.87--0.93
 & 0.91 & 0.88--0.93 \\
\bottomrule
\end{tabular}}
\caption{Adjusted KS prevalence analysis. Four estimators are compared: IPW, stabilised weight estimator (Stab$_{z}$), generalised raking (GR), and the proposed two-step approach that incorporates stabilised weights into generalised raking (Stab-rake).} \label{KS}
\end{table}

\section{Discussion}
\label{discussion}

In this paper, we evaluated the performance of stabilised weights and generalised raking for regression models in two-phase studies. Building on this framework, we also proposed combining stabilised weights with generalised raking. This combination leverages the strengths of both approaches: the stabilised weight estimator enhances efficiency by modifying the sampling weights to ignore part of the variation explained by variables in the model, while generalised raking improves efficiency by adjusting the sampling weights as a function of the phase-1 variables.

The central idea of stabilizing weights is to remove non-essential variation in the weights, which can otherwise lead to inflated variances of the regression parameter estimates. This principle is particularly relevant in settings such as dual-frame sampling \citep{Metcalf2009, Shepherd2023Multiwave}, where two samples are drawn from overlapping sampling frames and the raw weights are often highly variable. Scaling the weights in these cases improves precision by reducing non-essential variation in the combined weights. The cross-national surveys discussed by \cite{SKINNER2012} are a related problem. \cite{Scott2001Case} explored the case--control design in complex sampling and showed that efficiency gains could be achieved by assigning greater weight to cases in the weighted estimating equation. These efficiency improvements resulted from rescaling the weights as a function of the outcome variable $Y$. This approach is conceptually similar to the idea of weight stabilization.

In two-phase sampling, the design variables and their relationship with the sampling weights are typically well understood, which often leads to a highly informative sampling design with deliberate variation in the weights. In our simulation studies, when we considered the optimal design for design-based estimators \citep{Chen2022optimal}, the improvements from weight stabilization were limited. This suggests that much of the non-essential variation in the weights had already been addressed at the design stage.

However, results from the KS study underscore the synergistic gains achieved by combining generalised raking with stabilised weights in real-world applications. These gains in precision are obtained without the need to review additional charts (i.e., collect more phase-2 data) and without additional burdensome computation, suggesting that generalised raking with stabilised weights may be a useful strategy for improving estimation in two-phase designs.

Of note, our results extend beyond generalised linear models and can be generalised to semiparametric models. Similarly, the variation in weights can be reduced by multiplying the weights by a function of covariates. Additionally, our results also apply to other statistical functionals such as non-parametric causal estimands \citep{Chambers2022weight}.

The proposed methods will, in general, be less efficient than the semiparametric efficient estimators developed by \cite{Scott1997} and \cite{Tao2017Efficient}. However, the case--control design example shows that the stabilised weight estimator can achieve close to full efficiency. We note that, for logistic regression, the assumption of a correctly specified marginal mean model in the stabilised weight estimator is equivalent to assuming a correct parametric model for $Y$; the high efficiency of the stabilised weight estimator may therefore be a consequence of this. For other generalised linear models, the assumption of correct marginal mean specification is strictly weaker than that of a correct parametric model. Even when our method is less efficient than semiparametric maximum likelihood, it is easier to implement.

Future research is needed to investigate the robustness--efficiency trade-off for the stabilised weight estimators. While weight stabilization can improve precision, it is unclear whether the stabilised estimator is more robust to model misspecification compared to semiparametric efficient estimators.

\section*{ACKNOWLEDGMENTS}
This work was supported in part by the United States National Institutes of Health, grant numbers R37AI131771, U01AI069923, and U01AI069911. 

\bibliographystyle{abbrvnat}
\bibliography{references}  
\clearpage

\input{suppl}

\end{document}

%% file: suppl.tex
\begin{titlepage}
   \begin{center}
       \vspace*{1cm}

       \Huge{SUPPLEMENTARY MATERIAL\\}
       \vspace{3cm}
       \LARGE{
       Generalised raking and stabilised weights for regression modelling in two-phase samples}\\
       \vspace{4cm}
       \Large{Tong Chen$^{1,2}$, Joshua Slone$^{3}$, Gustavo Amorim$^{3}$, 
       Pamela A. Shaw$^{4}$, \\ Bryan E. Shepherd$^{3}$, Thomas Lumley$^{5}$} \\
       \vspace{3cm}

       \large{1. Clinical Epidemiology and Biostatistics Unit, \\Department of Paediatrics, University of Melbourne\\ 
       2. Clinical Epidemiology and Biostatistics Unit, Murdoch Children’s Research Institute\\
       3. Department of Biostatistics, Vanderbilt University\\
       4. Biostatistics Division, 
Kaiser Permanente Washington Health Research Institute \\
       5. Department of Statistics, University of Auckland}

       \vspace{3cm}
       
       \Large{\today}
            
   \end{center}
\end{titlepage}
\setcounter{section}{0}
\renewcommand{\thesection}{S\arabic{section}}
\renewcommand{\thetable}{S\arabic{table}} 
\setcounter{table}{0}

\section{Case-Control design with uniformly distributed exposure}

The continuous exposure $X$ was drawn from a uniform distribution, $X \sim \text{Unif}(0,1)$. The binary outcome $Y$ was simulated using the logistic regression model $\text{logit}(P(Y = 1 \mid X)) = \beta_0 + \beta_x X$, where $\text{logit}(x) = \text{log}(x/(1-x))$ and the intercept was set to $\beta_0 = -4$. The coefficient $\beta_x$ was varied to evaluate the performance of the proposed estimator across different effect sizes of exposure $X$. $1000$ case-control samples were simulated.

We considered the maximum likelihood estimator, the IPW estimator, and the stabilised weight estimators with $q(x,z)$ calculated following Equation (10) in the main manuscript.

\begin{table}[H]
\centering
\begin{tblr}
{
colspec={Q[]Q[]Q[]Q[]Q[]Q[]Q[]Q[]Q[]},
cell{1}{4}={c=2,}{halign=c,},
cell{1}{6}={c=2,}{halign=c,},
cell{1}{8}={c=2,}{halign=c,},
cell{3}{1}={r=2,}{valign=h,},
cell{5}{1}={r=2,}{valign=h,},
cell{7}{1}={r=2,}{valign=h,},
cell{9}{1}={r=2,}{valign=h,},
cell{3}{2}={r=2,}{valign=h,},
cell{5}{2}={r=2,}{valign=h,},
cell{7}{2}={r=2,}{valign=h,},
cell{9}{2}={r=2,}{valign=h,},
}
\toprule
& &  & MLE &  & Stab &  & IPW &  \\
\cmidrule[lr]{4-5}\cmidrule[lr]{6-7}\cmidrule[lr]{8-9}
$\Bar{n}$ & $\beta_x$ &  & empSE & RMSE & empSE & RMSE & empSE & RMSE \\
\midrule
360 & 0 & Intercept & 0.201 & 0.201 & 0.200 & 0.200 & 0.201 & 0.201 \\
360 & 0 & X         & 0.375 & 0.375 & 0.375 & 0.374 & 0.377 & 0.376 \\
463 & 0.5 & Intercept & 0.188 & 0.188 & 0.188 & 0.188 & 0.189 & 0.189 \\
463 & 0.5 & X         & 0.332 & 0.332 & 0.331 & 0.331 & 0.332 & 0.332 \\
607 & 1 & Intercept & 0.166 & 0.166 & 0.166 & 0.166 & 0.167 & 0.167 \\
607 & 1 & X         & 0.287 & 0.287 & 0.287 & 0.286 & 0.289 & 0.289 \\
809 & 1.5 & Intercept & 0.159 & 0.159 & 0.159 & 0.159 & 0.161 & 0.161 \\
809 & 1.5 & X         & 0.268 & 0.268 & 0.268 & 0.268 & 0.272 & 0.272 \\
\bottomrule
\end{tblr}
\caption{Empirical standard error (empSE) and root mean squared error (RMSE) for different estimators under varying values of $\beta_x$. $\Bar{n}$ represents the average phase-2 sample size over all simulation replicates. Three estimators are compared: MLE, IPW, and the stabilised weights estimator with $q(x,z)$ defined in Equation (10) in the main manuscript.}
\label{cc-unif}
\end{table}

\newpage

\section{General case--control sampling}
We next considered a general case--control study where the effect of exposure $X$ on the outcome $Y$ was confounded by $Z$. Unlike the case--control setting examined in the main text, the optimal stabilised weights are analytically intractable in this setting.

$1000$ datasets with $N = 10,000$ individuals were simulated and a case--control sample was drawn from each dataset. The variables were generated as follows: $Z \sim N(0,1), X \sim N(0.8 \times Z, 1).$ The binary outcome $Y$ was simulated using logistic regression $\text{logit}(P(Y = 1 \mid X, Z)) = \beta_0 + \beta_x X + \beta_z Z,$ where $\beta_0 = -4$. We implemented five estimators: the IPW estimator, generalised raking estimator (GR) where we calibrated on the influence function of $Z$, the proposed two-step approach that incorporated stabilised weights into generalised raking, and two stabilised weight estimators (Stab$_{xz}$ and Stab$_{z}$). The Stab$_{xz}$ estimator used $q(x,z)$ to stabilise on both $X$ and $Z$, whereas the Stab$_{z}$ estimator used $q(z)$.

The results are presented in Table~\ref{sim2}. Stab$_{xz}$ was more efficient than Stab$_{z}$ which only used $Z$ for stabilisation via $q(z)$. For estimation of $\beta_x$, the proposed two-step approach (Stab-rake) had efficiency comparable to $\text{Stab}_{z}$ and exceeded that of both the generalised raking and IPW estimators. For estimation of $\beta_z$, generalised raking was more efficient than IPW, $\text{Stab}z$, and $\text{Stab}{xz}$. The proposed two-step approach further improved upon generalised raking.

\begin{table}[h]
\centering
\resizebox{\textwidth}{!}{%
\begin{tblr}[
]                     
{                     
colspec={Q[]Q[]Q[]Q[]Q[]Q[]Q[]Q[]Q[]Q[]Q[]Q[]Q[]Q[]},
cell{1}{5}={c=2,}{halign=c,},
cell{1}{7}={c=2,}{halign=c,},
cell{1}{9}={c=2,}{halign=c,},
cell{1}{11}={c=2,}{halign=c,},
cell{1}{13}={c=2,}{halign=c,},
cell{3}{2}={r=2,}{valign=h,},
cell{5}{2}={r=2,}{valign=h,},
cell{7}{2}={r=2,}{valign=h,},
cell{9}{2}={r=2,}{valign=h,},
cell{11}{2}={r=2,}{valign=h,},
cell{13}{2}={r=2,}{valign=h,},
cell{15}{2}={r=2,}{valign=h,},
cell{17}{2}={r=2,}{valign=h,},
cell{19}{2}={r=2,}{valign=h,},
cell{3}{3}={r=2,}{valign=h,},
cell{5}{3}={r=2,}{valign=h,},
cell{7}{3}={r=2,}{valign=h,},
cell{9}{3}={r=2,}{valign=h,},
cell{11}{3}={r=2,}{valign=h,},
cell{13}{3}={r=2,}{valign=h,},
cell{15}{3}={r=2,}{valign=h,},
cell{17}{3}={r=2,}{valign=h,},
cell{19}{3}={r=2,}{valign=h,},
cell{3}{1}={r=6,}{valign=h,},
cell{9}{1}={r=6,}{valign=h,},
cell{15}{1}={r=6,}{valign=h,},
}                     
\toprule
&  &  &  & Stab$_{xz}$ &  & Stab$_z$ &  & Stab-rake &  & IPW &  & GR &  \\ 
\cmidrule[lr]{5-6}\cmidrule[lr]{7-8}\cmidrule[lr]{9-10}\cmidrule[lr]{11-12}\cmidrule[lr]{13-14}
$\beta_z$ & $\beta_x$ & $\bar{n}$ & & empSE & RMSE & empSE & RMSE & empSE & RMSE & empSE & RMSE & empSE & RMSE \\ 
\midrule
0.1 & 0.1 & 367 & X & 0.106 & 0.106 & 0.110 & 0.110 & 0.110 & 0.110 & 0.110 & 0.110 & 0.110 & 0.110 \\
0.1 & 0.1 & 367 & Z & 0.134 & 0.134 & 0.137 & 0.137 & 0.113 & 0.113 & 0.137 & 0.137 & 0.113 & 0.113 \\
0.1 & 0.5 & 453 & X & 0.107 & 0.108 & 0.118 & 0.119 & 0.118 & 0.119 & 0.119 & 0.120 & 0.118 & 0.120 \\
0.1 & 0.5 & 453 & Z & 0.125 & 0.125 & 0.137 & 0.137 & 0.111 & 0.111 & 0.137 & 0.137 & 0.111 & 0.111 \\
0.1 & 1.0 & 763 & X & 0.101 & 0.101 & 0.128 & 0.130 & 0.128 & 0.130 & 0.130 & 0.132 & 0.130 & 0.132 \\
0.1 & 1.0 & 763 & Z & 0.112 & 0.112 & 0.132 & 0.133 & 0.112 & 0.114 & 0.135 & 0.135 & 0.112 & 0.113 \\
0.5 & 0.1 & 422 & X & 0.101 & 0.101 & 0.108 & 0.108 & 0.108 & 0.108 & 0.109 & 0.109 & 0.109 & 0.109 \\
0.5 & 0.1 & 422 & Z & 0.132 & 0.132 & 0.144 & 0.144 & 0.109 & 0.109 & 0.146 & 0.146 & 0.110 & 0.110 \\
0.5 & 0.5 & 576 & X & 0.095 & 0.095 & 0.113 & 0.114 & 0.112 & 0.113 & 0.115 & 0.116 & 0.115 & 0.116 \\
0.5 & 0.5 & 576 & Z & 0.121 & 0.121 & 0.136 & 0.135 & 0.099 & 0.099 & 0.141 & 0.141 & 0.101 & 0.101 \\
0.5 & 1.0 & 999 & X & 0.092 & 0.092 & 0.115 & 0.116 & 0.116 & 0.116 & 0.119 & 0.120 & 0.119 & 0.120 \\
0.5 & 1.0 & 999 & Z & 0.108 & 0.108 & 0.126 & 0.126 & 0.095 & 0.096 & 0.135 & 0.135 & 0.096 & 0.096 \\
1.0 & 0.1 & 606 & X & 0.093 & 0.093 & 0.107 & 0.107 & 0.107 & 0.107 & 0.112 & 0.112 & 0.111 & 0.111 \\
1.0 & 0.1 & 606 & Z & 0.130 & 0.130 & 0.152 & 0.154 & 0.107 & 0.107 & 0.163 & 0.165 & 0.112 & 0.113 \\
1.0 & 0.5 & 872 & X & 0.085 & 0.085 & 0.101 & 0.101 & 0.101 & 0.101 & 0.108 & 0.108 & 0.107 & 0.108 \\
1.0 & 0.5 & 872 & Z & 0.112 & 0.112 & 0.130 & 0.130 & 0.089 & 0.089 & 0.146 & 0.147 & 0.097 & 0.097 \\
1.0 & 1.0 & 1418 & X & 0.082 & 0.083 & 0.098 & 0.099 & 0.098 & 0.099 & 0.104 & 0.104 & 0.104 & 0.104 \\
1.0 & 1.0 & 1418 & Z & 0.102 & 0.102 & 0.117 & 0.116 & 0.088 & 0.088 & 0.129 & 0.129 & 0.092 & 0.092 \\
\bottomrule
\end{tblr}}
\caption{Empirical standard error (empSE) and root mean squared error (RMSE) of $X$ and $Z$ for different estimators under varying values of $\beta_x$ and $\beta_z$. $\Bar{n}$ represents the average phase-2 sample size over all simulation replicates.}
\label{sim2}
\end{table}

\newpage

\section{General two-phase sampling with continuous outcomes}

\subsection{Stratification on Z with optimal design}

We considered the same setting as in the main manuscript, with the only difference being that the optimal design for design-based subsampling was applied. 1000 datasets with $N = 6000$ individuals were generated. The variables were generated as follows: $W_1 \sim N(0,1)$ and $W_2 \sim \text{Bernoulli}(0.6)$. The phase-2 variable $X$ was defined as $X = 0.3 + W_1 + W_2 + \epsilon_X$ with $\epsilon_X \sim N(0,1)$, and the phase-1 variable $Z$ was defined as $Z = W_1 + 0.8 \times W_2 + \epsilon_Z$ with $\epsilon_Z \sim N(0,1)$. An additional variable $A$ was defined as $A = X + \epsilon_A$, with $\epsilon_A \sim N(0,1)$ and was used in generalised raking. The outcome was generated from the linear model $Y = \beta_x X + \beta_z Z + \epsilon_Y$, where $\epsilon_Y \sim N(0,\sigma^2)$. We set $\beta_x = \beta_z$ and varied their values, as well as the residual variance $\sigma^2$. Stratification was based on the 90th percentile of $Z$, and the optimal design for the design-based estimators, obtained by applying Neyman allocation based on influence functions, was considered, and 600 individuals were selected for the phase-2 sample.

We considered the same set of estimators, including IPW, generalised raking calibrated on the influence function of $Z$, the proposed two step approach incorporating stabilised weights into generalised raking, and two stabilised weight estimators, $\text{Stab}_{xz}$ and $\text{Stab}_z$. In this setting, weight stabilisation did not lead to further efficiency gains, suggesting that variation in the weights was limited under the optimal design.

\begin{table}[H]
\centering
\resizebox{\textwidth}{!}{%
\begin{tblr}
{
colspec={Q[]Q[]Q[]Q[]Q[]Q[]Q[]Q[]Q[]Q[]Q[]Q[]Q[]},
cell{1}{4}={c=2,}{halign=c,},
cell{1}{6}={c=2,}{halign=c,},
cell{1}{8}={c=2,}{halign=c,},
cell{1}{10}={c=2,}{halign=c,},
cell{1}{12}={c=2,}{halign=c,},
cell{3}{2}={r=2,}{valign=h,},
cell{5}{2}={r=2,}{valign=h,},
cell{7}{2}={r=2,}{valign=h,},
cell{9}{2}={r=2,}{valign=h,},
cell{11}{2}={r=2,}{valign=h,},
cell{13}{2}={r=2,}{valign=h,},
cell{15}{2}={r=2,}{valign=h,},
cell{17}{2}={r=2,}{valign=h,},
cell{3}{1}={r=8,}{valign=h,},
cell{11}{1}={r=8,}{valign=h,},
}
\toprule
&  &  & IPW &  & Stab$_{xz}$ &  & Stab$_z$ &  & Stab-rake &  & GR &  \\
\cmidrule[lr]{4-5}\cmidrule[lr]{6-7}\cmidrule[lr]{8-9}\cmidrule[lr]{10-11}\cmidrule[lr]{12-13}
$\sigma^2$ & $\beta_x \& \beta_z$ &  & empSE & RMSE & empSE & RMSE & empSE & RMSE & empSE & RMSE & empSE & RMSE \\
\midrule
0.5 & 0.1 & X & 0.022 & 0.022 & 0.022 & 0.022 & 0.022 & 0.022 & 0.015 & 0.015 & 0.015 & 0.015 \\
0.5 & 0.1 & Z & 0.024 & 0.024 & 0.024 & 0.024 & 0.024 & 0.024 & 0.011 & 0.011 & 0.011 & 0.011 \\
0.5 & 0.5 & X & 0.024 & 0.024 & 0.024 & 0.024 & 0.024 & 0.024 & 0.018 & 0.018 & 0.018 & 0.018 \\
0.5 & 0.5 & Z & 0.024 & 0.024 & 0.024 & 0.024 & 0.024 & 0.024 & 0.016 & 0.016 & 0.016 & 0.016 \\
0.5 & 1.0 & X & 0.023 & 0.023 & 0.023 & 0.023 & 0.023 & 0.023 & 0.020 & 0.020 & 0.020 & 0.020 \\
0.5 & 1.0 & Z & 0.024 & 0.024 & 0.024 & 0.024 & 0.024 & 0.024 & 0.021 & 0.021 & 0.021 & 0.021 \\
0.5 & 1.5 & X & 0.023 & 0.023 & 0.023 & 0.023 & 0.023 & 0.023 & 0.021 & 0.021 & 0.021 & 0.021 \\
0.5 & 1.5 & Z & 0.023 & 0.023 & 0.023 & 0.023 & 0.023 & 0.023 & 0.021 & 0.021 & 0.021 & 0.021 \\
1.0 & 0.1 & X & 0.032 & 0.032 & 0.032 & 0.032 & 0.032 & 0.032 & 0.022 & 0.022 & 0.022 & 0.022 \\
1.0 & 0.1 & Z & 0.033 & 0.033 & 0.034 & 0.034 & 0.034 & 0.034 & 0.015 & 0.015 & 0.015 & 0.015 \\
1.0 & 0.5 & X & 0.033 & 0.033 & 0.033 & 0.033 & 0.033 & 0.033 & 0.024 & 0.024 & 0.024 & 0.024 \\
1.0 & 0.5 & Z & 0.032 & 0.032 & 0.032 & 0.032 & 0.032 & 0.032 & 0.020 & 0.020 & 0.020 & 0.020 \\
1.0 & 1.0 & X & 0.033 & 0.033 & 0.033 & 0.033 & 0.033 & 0.033 & 0.026 & 0.026 & 0.026 & 0.026 \\
1.0 & 1.0 & Z & 0.035 & 0.035 & 0.035 & 0.035 & 0.035 & 0.035 & 0.026 & 0.026 & 0.026 & 0.026 \\
1.0 & 1.5 & X & 0.032 & 0.033 & 0.033 & 0.033 & 0.033 & 0.033 & 0.028 & 0.028 & 0.028 & 0.028 \\
1.0 & 1.5 & Z & 0.034 & 0.034 & 0.034 & 0.034 & 0.034 & 0.034 & 0.030 & 0.030 & 0.029 & 0.029 \\
\bottomrule
\end{tblr}}
\caption{Empirical standard error (empSE) and root mean squared error (RMSE) of $X$ and $Z$ for different estimators under varying values of $\beta_x$, $\beta_z$ and $\sigma^2$.}
\end{table}

\newpage
\subsubsection{Stratification on $A$ and $Z$ with balanced design}

We next considered two-phase designs in which stratification was based on both $A$ and $Z$. As before, 1000 datasets with $N = 6000$ individuals were generated. The variables were generated as follows: $W_1 \sim N(0,1)$ and $W_2 \sim \text{Bernoulli}(0.6)$. The phase-1 variable $Z$ was generated from a Bernoulli distribution with $\text{logit}(P(Z=1 \mid W_1, W_2)) = 0.3 + W_1 + W_2$. The phase-2 variable $X$ was defined as $X = W_1 + 0.8 \times W_2 + Z + \epsilon_X$, with $\epsilon_X \sim N(0,1)$. An additional variable $A$ was defined as $A = X + \epsilon_A$, with $\epsilon_A \sim N(0,1)$, and was used in both stratification and generalised raking. The outcome was generated from the linear model $Y = \beta_x X + \beta_z Z + \epsilon_Y$, where $\epsilon_Y \sim N(0, \sigma^2)$. We set $\beta_x = \beta_z$ and varied their values, as well as the residual variance $\sigma^2$. Stratification was based on the cross-classification of $A$ and $Z$, where $A$ was divided at its 30th and 70th percentiles, resulting in 6 strata. A balanced design was applied, and 600 individuals were selected for the phase-2 sample.

\begin{table}[htbp]
\resizebox{\textwidth}{!}{%
\centering
\begin{tblr}
{
colspec={Q[]Q[]Q[]Q[]Q[]Q[]Q[]Q[]Q[]Q[]Q[]Q[]Q[]},
cell{1}{4}={c=2,}{halign=c,},
cell{1}{6}={c=2,}{halign=c,},
cell{1}{8}={c=2,}{halign=c,},
cell{1}{10}={c=2,}{halign=c,},
cell{1}{12}={c=2,}{halign=c,},
cell{3}{2}={r=2,}{valign=h,},
cell{5}{2}={r=2,}{valign=h,},
cell{7}{2}={r=2,}{valign=h,},
cell{9}{2}={r=2,}{valign=h,},
cell{11}{2}={r=2,}{valign=h,},
cell{13}{2}={r=2,}{valign=h,},
cell{15}{2}={r=2,}{valign=h,},
cell{17}{2}={r=2,}{valign=h,},
cell{19}{2}={r=2,}{valign=h,},
cell{3}{1}={r=6,}{valign=h,},
cell{9}{1}={r=6,}{valign=h,},
cell{15}{1}={r=6,}{valign=h,},
}
\toprule
&  &  & IPW &  & Stab$_{xz}$ &  & Stab$_z$ &  & Stab-rake &  & GR &  \\
\cmidrule[lr]{4-5}\cmidrule[lr]{6-7}\cmidrule[lr]{8-9}\cmidrule[lr]{10-11}\cmidrule[lr]{12-13}
$\sigma^2$ & $\beta_x \& \beta_z$ &  & empSE & RMSE & empSE & RMSE & empSE & RMSE & empSE & RMSE & empSE & RMSE \\
\midrule
0.5 & 0.1 & X & 0.022 & 0.022 & 0.022 & 0.022 & 0.022 & 0.022 & 0.016 & 0.016 & 0.016 & 0.016 \\
0.5 & 0.1 & Z & 0.072 & 0.072 & 0.071 & 0.071 & 0.071 & 0.071 & 0.036 & 0.036 & 0.036 & 0.036 \\
0.5 & 0.5 & X & 0.022 & 0.022 & 0.022 & 0.022 & 0.022 & 0.022 & 0.017 & 0.017 & 0.017 & 0.017 \\
0.5 & 0.5 & Z & 0.072 & 0.072 & 0.071 & 0.071 & 0.071 & 0.071 & 0.046 & 0.046 & 0.046 & 0.046 \\
0.5 & 1 & X & 0.022 & 0.022 & 0.022 & 0.022 & 0.022 & 0.022 & 0.019 & 0.019 & 0.019 & 0.019 \\
0.5 & 1 & Z & 0.072 & 0.072 & 0.071 & 0.071 & 0.071 & 0.071 & 0.058 & 0.058 & 0.059 & 0.059 \\
1.0 & 0.1 & X & 0.031 & 0.031 & 0.031 & 0.031 & 0.031 & 0.031 & 0.022 & 0.022 & 0.022 & 0.022 \\
1.0 & 0.1 & Z & 0.102 & 0.102 & 0.101 & 0.101 & 0.101 & 0.101 & 0.051 & 0.051 & 0.050 & 0.050 \\
1.0 & 0.5 & X & 0.031 & 0.031 & 0.031 & 0.031 & 0.031 & 0.031 & 0.023 & 0.023 & 0.023 & 0.023 \\
1.0 & 0.5 & Z & 0.102 & 0.102 & 0.101 & 0.101 & 0.101 & 0.101 & 0.059 & 0.059 & 0.059 & 0.059 \\
1.0 & 1 & X & 0.031 & 0.031 & 0.031 & 0.031 & 0.031 & 0.031 & 0.025 & 0.025 & 0.025 & 0.025 \\
1.0 & 1 & Z & 0.102 & 0.102 & 0.101 & 0.101 & 0.101 & 0.101 & 0.074 & 0.074 & 0.074 & 0.074 \\
2.0 & 0.1 & X & 0.044 & 0.044 & 0.044 & 0.044 & 0.043 & 0.043 & 0.031 & 0.031 & 0.031 & 0.031 \\
2.0 & 0.1 & Z & 0.145 & 0.145 & 0.142 & 0.142 & 0.142 & 0.142 & 0.071 & 0.071 & 0.071 & 0.071 \\
2.0 & 0.5 & X & 0.044 & 0.044 & 0.044 & 0.044 & 0.043 & 0.043 & 0.032 & 0.032 & 0.032 & 0.032 \\
2.0 & 0.5 & Z & 0.145 & 0.145 & 0.142 & 0.142 & 0.142 & 0.142 & 0.078 & 0.078 & 0.077 & 0.077 \\
2.0 & 1 & X & 0.044 & 0.044 & 0.044 & 0.044 & 0.043 & 0.043 & 0.033 & 0.033 & 0.033 & 0.033 \\
2.0 & 1 & Z & 0.145 & 0.145 & 0.142 & 0.142 & 0.142 & 0.142 & 0.092 & 0.092 & 0.093 & 0.093 \\
\bottomrule
\end{tblr}}
\caption{Empirical standard error (empSE) and root mean squared error (RMSE) of $X$ and $Z$ for different estimators under varying values of $\beta_x$, $\beta_z$ and $\sigma^2$. Five estimators are compared: IPW, generalized raking (GR), the proposed two-step approach that incorporates stabilized weights into generalized raking (Stab-rake), and two stabilized weight estimators (Stab$_{xz}$ and Stab$_{z}$). The Stab$_{xz}$ estimator uses $q(x,z)$ as defined in Equation~(4) in the main manuscript, while the Stab$_{z}$ estimator uses $q(z)$ as defined in Equation (5) in the main manuscript.}
\label{sim4}
\end{table}

The results are presented in Table~\ref{sim4}. The proposed two-step approach (Stab-rake) was as efficient as generalised raking estimator in all settings. The stabilised weight estimators (Stab$_{xz}$ and Stab$_{z}$) performed similarly to the IPW estimator. These results suggested that the non-essential variation in the weights was small under these two-phase designs, so the weights stabilisation contributed little to improving efficiency. Consistently when the optimal design for design-based estimators was considered (Section~\ref{az-opt}), no additional efficiency gain was observed through weight stabilisation.

\newpage

\subsection{Stratification on A and Z with optimal design} \label{az-opt}

We considered the same setting as in the main manuscript, with the only difference being that the optimal design for design-based subsampling was applied. As before, 1000 datasets with $N = 6000$ individuals were generated. The variables were generated as follows: $W_1 \sim N(0,1)$ and $W_2 \sim \text{Bernoulli}(0.6)$. The phase-1 variable $Z$ was generated from a Bernoulli distribution with $\text{logit}(P(Z=1 \mid W_1, W_2)) = 0.3 + W_1 + W_2$. The phase-2 variable $X$ was defined as $X = W_1 + 0.8 \times W_2 + Z + \epsilon_X$, with $\epsilon_X \sim N(0,1)$. An additional variable $A$ was defined as $A = X + \epsilon_A$, with $\epsilon_A \sim N(0,1)$, and was used in both stratification and generalised raking. The outcome was generated from the linear model $Y = \beta_x X + \beta_z Z + \epsilon_Y$, where $\epsilon_Y \sim N(0, \sigma^2)$. We set $\beta_x = \beta_z$ and varied their value, as well as the residual variance $\sigma^2$. Stratification was based on the cross-classification of $A$ and $Z$, where $A$ was divided at its 30th and 70th percentiles, resulting in 6 strata. 600 individuals were selected for the phase-2 sample.

\begin{table}[H]
\centering
\resizebox{\textwidth}{!}{%
\begin{tblr}
{
colspec={Q[]Q[]Q[]Q[]Q[]Q[]Q[]Q[]Q[]Q[]Q[]Q[]Q[]},
cell{1}{4}={c=2,}{halign=c,},
cell{1}{6}={c=2,}{halign=c,},
cell{1}{8}={c=2,}{halign=c,},
cell{1}{10}={c=2,}{halign=c,},
cell{1}{12}={c=2,}{halign=c,},
cell{3}{2}={r=2,}{valign=h,},
cell{5}{2}={r=2,}{valign=h,},
cell{7}{2}={r=2,}{valign=h,},
cell{9}{2}={r=2,}{valign=h,},
cell{11}{2}={r=2,}{valign=h,},
cell{13}{2}={r=2,}{valign=h,},
cell{15}{2}={r=2,}{valign=h,},
cell{17}{2}={r=2,}{valign=h,},
cell{19}{2}={r=2,}{valign=h,},
cell{3}{1}={r=6,}{valign=h,},
cell{9}{1}={r=6,}{valign=h,},
cell{15}{1}={r=6,}{valign=h,},
}
\toprule
&  &  & IPW &  & Stab$_{xz}$ &  & Stab$_z$ &  & Stab-rake &  & GR &  \\
\cmidrule[lr]{4-5}\cmidrule[lr]{6-7}\cmidrule[lr]{8-9}\cmidrule[lr]{10-11}\cmidrule[lr]{12-13}
$\sigma^2$ & $\beta_x \& \beta_z$ &  & empSE & RMSE & empSE & RMSE & empSE & RMSE & empSE & RMSE & empSE & RMSE \\
\midrule
0.5 & 0.1 & X & 0.021 & 0.021 & 0.021 & 0.021 & 0.021 & 0.021 & 0.014 & 0.014 & 0.014 & 0.014 \\
0.5 & 0.1 & Z & 0.079 & 0.079 & 0.079 & 0.079 & 0.079 & 0.079 & 0.035 & 0.035 & 0.035 & 0.035 \\
0.5 & 0.5 & X & 0.020 & 0.020 & 0.020 & 0.020 & 0.020 & 0.020 & 0.015 & 0.015 & 0.015 & 0.015 \\
0.5 & 0.5 & Z & 0.076 & 0.076 & 0.076 & 0.076 & 0.076 & 0.076 & 0.049 & 0.049 & 0.049 & 0.049 \\
0.5 & 1.0 & X & 0.020 & 0.020 & 0.021 & 0.021 & 0.021 & 0.021 & 0.017 & 0.017 & 0.017 & 0.017 \\
0.5 & 1.0 & Z & 0.079 & 0.079 & 0.079 & 0.079 & 0.079 & 0.079 & 0.064 & 0.064 & 0.064 & 0.064 \\
1.0 & 0.1 & X & 0.029 & 0.029 & 0.029 & 0.029 & 0.029 & 0.029 & 0.020 & 0.020 & 0.020 & 0.020 \\
1.0 & 0.1 & Z & 0.110 & 0.110 & 0.110 & 0.110 & 0.110 & 0.110 & 0.051 & 0.051 & 0.050 & 0.050 \\
1.0 & 0.5 & X & 0.029 & 0.029 & 0.029 & 0.029 & 0.029 & 0.029 & 0.021 & 0.021 & 0.020 & 0.020 \\
1.0 & 0.5 & Z & 0.115 & 0.115 & 0.115 & 0.115 & 0.115 & 0.115 & 0.061 & 0.061 & 0.061 & 0.061 \\
1.0 & 1.0 & X & 0.029 & 0.029 & 0.030 & 0.030 & 0.029 & 0.030 & 0.023 & 0.023 & 0.023 & 0.023 \\
1.0 & 1.0 & Z & 0.110 & 0.110 & 0.111 & 0.111 & 0.110 & 0.110 & 0.078 & 0.078 & 0.078 & 0.078 \\
2.0 & 0.1 & X & 0.041 & 0.041 & 0.042 & 0.042 & 0.041 & 0.041 & 0.027 & 0.027 & 0.027 & 0.027 \\
2.0 & 0.1 & Z & 0.152 & 0.153 & 0.152 & 0.152 & 0.153 & 0.153 & 0.066 & 0.067 & 0.066 & 0.066 \\
2.0 & 0.5 & X & 0.041 & 0.041 & 0.041 & 0.041 & 0.041 & 0.041 & 0.027 & 0.027 & 0.027 & 0.027 \\
2.0 & 0.5 & Z & 0.156 & 0.156 & 0.156 & 0.156 & 0.156 & 0.155 & 0.074 & 0.074 & 0.074 & 0.074 \\
2.0 & 1.0 & X & 0.040 & 0.040 & 0.041 & 0.041 & 0.041 & 0.041 & 0.030 & 0.030 & 0.030 & 0.030 \\
2.0 & 1.0 & Z & 0.151 & 0.151 & 0.152 & 0.152 & 0.151 & 0.151 & 0.098 & 0.098 & 0.098 & 0.098 \\
\bottomrule
\end{tblr}}
\caption{Empirical standard error (empSE) and root mean squared error (RMSE) of $X$ and $Z$ for different estimators under varying values of $\beta_x$, $\beta_z$ and $\sigma^2$.}
\end{table}

\newpage
\section{Binary Outcome}
We examined the performance of the proposed estimators for binary outcomes under optimal designs for design-based estimators. 1000 datasets with $N = 10,000$ individuals were generated. The variables were generated as follows: $W_1 \sim N(0,1)$ and $W_2 \sim \text{Bernoulli}(0.6)$. The phase-2 variable $X$ was defined as $X = 0.3 + W_1 + W_2 + \epsilon_X$, with $\epsilon_X \sim N(0,1)$, and the phase-1 covariate $Z$ was defined as $Z = W_1 + 0.8 \times W_2 + \epsilon_Z$, with $\epsilon_Z \sim N(0,1)$. An additional variable $A$ was defined as $A = X + \epsilon_A$, with $\epsilon_A \sim N(0,1)$ and was used in generalised raking. A binary outcome $Y$ was generated from $\text{logit}(P(Y=1 \mid X,Z)) = -4 + \beta_x X + \beta_z Z$. Stratification was based on the cross-classification of $A$ and $Z$, where $A$ was divided at its 30th and 70th percentiles and $Z$ at its 60th percentile, resulting in 6 strata in total. $2000$ individuals were selected for the phase-2 sample.

\begin{table}[H]
\centering
\resizebox{\textwidth}{!}{%
\begin{tblr}
{
colspec={Q[]Q[]Q[]Q[]Q[]Q[]Q[]Q[]Q[]Q[]},
cell{1}{3}={c=2,}{halign=c,},
cell{1}{5}={c=2,}{halign=c,},
cell{1}{7}={c=2,}{halign=c,},
cell{1}{9}={c=2,}{halign=c,},
cell{3}{1}={r=2,}{valign=h,},
cell{5}{1}={r=2,}{valign=h,},
cell{7}{1}={r=2,}{valign=h,},
cell{9}{1}={r=2,}{valign=h,},
cell{11}{1}={r=2,}{valign=h,},
cell{13}{1}={r=2,}{valign=h,},
cell{15}{1}={r=2,}{valign=h,},
}
\toprule
&  & IPW &  & Stab$_z$ &  & Stab-rake &  & GR &  \\
\cmidrule[lr]{3-4}\cmidrule[lr]{5-6}\cmidrule[lr]{7-8}\cmidrule[lr]{9-10}
$\beta_x \& \beta_z$ &  & empSE & RMSE & empSE & RMSE & empSE & RMSE & empSE & RMSE \\
\midrule
-1.0 & X & 0.106 & 0.107 & 0.102 & 0.102 & 0.086 & 0.086 & 0.089 & 0.089 \\
-1.0 & Z & 0.117 & 0.118 & 0.111 & 0.116 & 0.082 & 0.090 & 0.081 & 0.082 \\
0.1  & X & 0.125 & 0.125 & 0.125 & 0.125 & 0.100 & 0.100 & 0.101 & 0.101 \\
0.1  & Z & 0.142 & 0.142 & 0.135 & 0.136 & 0.080 & 0.081 & 0.081 & 0.081 \\
0.5  & X & 0.074 & 0.074 & 0.071 & 0.071 & 0.056 & 0.056 & 0.059 & 0.059 \\
0.5  & Z & 0.086 & 0.086 & 0.082 & 0.084 & 0.051 & 0.055 & 0.051 & 0.051 \\
1.0  & X & 0.069 & 0.069 & 0.067 & 0.067 & 0.059 & 0.059 & 0.060 & 0.060 \\
1.0  & Z & 0.077 & 0.076 & 0.076 & 0.080 & 0.045 & 0.051 & 0.044 & 0.044 \\
1.5  & X & 0.088 & 0.089 & 0.088 & 0.088 & 0.081 & 0.081 & 0.081 & 0.082 \\
1.5  & Z & 0.101 & 0.102 & 0.103 & 0.104 & 0.062 & 0.065 & 0.062 & 0.062 \\
2.0  & X & 0.122 & 0.122 & 0.122 & 0.122 & 0.115 & 0.115 & 0.115 & 0.115 \\
2.0  & Z & 0.133 & 0.135 & 0.135 & 0.135 & 0.095 & 0.096 & 0.094 & 0.095 \\
3.0  & X & 0.183 & 0.185 & 0.185 & 0.185 & 0.181 & 0.181 & 0.179 & 0.181 \\
3.0  & Z & 0.191 & 0.192 & 0.194 & 0.195 & 0.168 & 0.169 & 0.165 & 0.167 \\
\bottomrule
\end{tblr}}
\caption{Empirical standard error (empSE) and root mean squared error (RMSE) of $X$ and $Z$ for different estimators under varying values of $\delta$. Four estimators are compared: IPW, generalised raking (GR), the proposed two-step approach that incorporates stabilised weights into generalised raking (Stab-rake), and the stabilised weight estimator (Stab$_z$). The Stab$_z$ estimator uses $q(z)$ as defined in Equation (5) in the main manuscript.}
\label{sim6}
\end{table}

The results are presented in Table~\ref{sim6}. The results indicated that the stabilised weight estimator (Stab$_z$) was as efficient as the IPW estimator, and the proposed two-step estimator was as efficient as the generalised raking estimator. These findings suggested that the variation in the weights was small, so there was little room for efficiency improvement through weight stabilisation.
